\documentclass[aps,prl,twocolumn,superscriptaddress]{revtex4-1}
\usepackage{graphicx}
\begin{document}

\title{Crystal plasticity finite element simulation of lattice rotation and x-ray diffraction during laser shock-compression of Tantalum}


\author{P. Avraam}
\email[]{philip.avraam@awe.co.uk}
\affiliation{Materials Physics Group, AWE, Aldermaston, Reading RG7 4PR, 
United Kingdom}
\author{D. McGonegle}
\affiliation{Materials Physics Group, AWE, Aldermaston, Reading RG7 4PR, 
United Kingdom}
\author{P. G. Heighway}
\affiliation{Department of Physics, Clarendon Laboratory, University of 
Oxford, Parks Road, Oxford, OX1 3PU, United Kingdom}
\author{C. E. Wehrenberg}
\affiliation{Lawrence Livermore National Laboratory, Livermore, California 94550, USA}
\author{E. Floyd}
\affiliation{Materials Physics Group, AWE, Aldermaston, Reading RG7 4PR, 
United Kingdom}
\author{A. Comley}
\affiliation{Materials Physics Group, AWE, Aldermaston, Reading RG7 4PR, 
United Kingdom}
\author{J. M. Foster}
\affiliation{Materials Physics Group, AWE, Aldermaston, Reading RG7 4PR, 
United Kingdom}
\author{J. Turner}
\affiliation{Materials Physics Group, AWE, Aldermaston, Reading RG7 4PR, 
United Kingdom}
\author{S. Case}
\affiliation{Materials Physics Group, AWE, Aldermaston, Reading RG7 4PR, 
United Kingdom}
\author{J. S. Wark}
\affiliation{Department of Physics, Clarendon Laboratory, University of 
Oxford, Parks Road, Oxford, OX1 3PU, United Kingdom}


\date{\today}


\begin{abstract}
Wehrenberg {\it et al} [{Nature 550 496 (2017)}] used ultrafast {\it in situ} x-ray diffraction
at the LCLS x-ray free-electron laser facility
to measure large lattice rotations resulting from slip and deformation
twinning in shock-compressed laser-driven [110] fibre textured tantalum polycrystal.
We employ a crystal plasticity finite 
element method model, with slip kinetics based closely on the isotropic dislocation-based Livermore Multiscale Model [Barton {\it et al.}, J. Appl. Phys. 109 (2011)], to analyse this experiment.
We elucidate the link between the degree of lattice rotation and the kinetics of plasticity, demonstrating that a transition occurs at shock pressures of $\sim$27~GPa, 
between a regime of relatively slow kinetics, resulting in a balanced pattern of
slip system activation and therefore relatively small net lattice rotation, and a regime
of fast kinetics, due to the onset of nucleation, resulting in a lop-sided pattern of deformation-system activation and therefore large net lattice rotations. 
We demonstrate a good fit between this model and experimental x-ray diffraction data of lattice rotation, and show that this data is constraining of deformation kinetics.
\end{abstract}

\maketitle


When a crystal is uniaxially shock compressed beyond the elastic limit,
plastic deformation mechanisms are activated. 
The high rates of plastic strain observed at
the shock front cannot be mediated by the motion of preexisting 
dislocations -- dislocation density must grow rapidly
during deformation at the shock front, or other plastic deformation mechanisms, 
such as deformation twinning, must occur. 
This deformation activity is anisotropic, occuring 
predominantly in specific characteristic crystallographic planes and directions
within the single-crystal grains that make up the polycrystal, 
resulting in lattice rotation (texture evolution). 

An understanding of the plastic 
deformation mechanisms 
and their kinetics spanning the relevant length-scales has long been sought.
An insight into the underlying physics of the resistance of materials to plastic flow at ultra-high strain rates and pressures is of direct relevance to a range of phenomena, including planetary impact \cite{Asphaug2006,Canup2001,Senft2007} and inertial confinement fusion \cite{Rudd2010}.  In this latter case, of particular interest is how material strength reduces the growth rate of hydrodynamic instabilities in a solid, with integrated experiments interrogating the growth of the Rayleigh-Taylor (RT) instability at strain rates of order 10$^6$-10$^{7}$s$^{-1}$  \cite{Park2015,Remington2019} showing good agreement with certain continuum-based approaches such as the Livermore multiscale strength (LMS) model \cite{doi:10.1063/1.3553718,Remington2019}.

Whilst such integrated RT experiments give some information on overall strength under these extreme conditions, no insight at the lattice level related to specific plasticity mechanisms can be directly ascertained.  However, discernment of plasticity-related activity at such a scale can be obtained within experiments that utilise ultrafast {\it in situ} 
X-ray diffraction (XRD) to study laser driven material undergoing
rapid plastic relaxation \cite{PhysRevB.92.104305,SuggitEtAl_2012, Milathianaki2013}.  
In particular, the seminal work of Wehrenberg {\it et al} \cite{Wehrenberg2017},
using the LCLS x-ray free-electron laser (XFEL) facility,
exploited initially highly textured polycrystalline 
samples to measure large lattice rotations due to slip and deformation twinning
during shock loading. The degree of rotation seems to imply slip (or twinning) on mainly one system, but why this is the case in the experiment has not been understood until this work. 
They compared their results with both a simple zero-dimensional single-slip analysis based on a the `Schmid' kinematic framework 
\cite{Schmid1926} -- with moderate success -- and with the results of large-scale classical molecular dynamics (MD) simulations of single crystal Tantalum.  However, it is now understood that the Schmid framework is fundamentally unsuited to describing lattice rotation under uniaxial compression \cite{Heighway2021}, and while large-scale classical molecular dynamics (MD) simulations
\cite{PhysRevB.88.134101,Heighway2021,PhysRevB.88.104105,TRAMONTINA20149,Zepeda-Ruiz_2017,Zepeda-Ruiz2021}, 
have yielded significant insight over the past decade, they are limited in the sizes and timescales that can be reached, due to computational cost, which restrict their utility in addressing
questions of texture evolution in polycrystals.

In contrast the crystal plasticity finite element (CPFE) method is a continuum
technique that overcomes these limitations. 
It provides a way to model polycrystals by explicitly 
accounting for the anisotropic plasticity and elasticity of single 
crystal grains, and grain-grain interaction, going beyond the widely used
mean-field homogenisation assumptions 
(e.g. the Voigt \cite{doi:10.1063/1.4874656} and Reuss \cite{doi:10.1063/1.4953028} 
assumptions).
The CPFE method keeps track of the lattice deformation and rotation, 
which can be used to calculate XRD patterns for comparison with experimental 
patterns.  The CPFE approach can incorporate and extend models such as the LMS model, which themselves rely for their construction on input from dislocation dynamics (DD), MD, and quantum mechanical simulations, affording a bridge between atomistic and meso-scale simulations.

It is in the above context that we present here results  from CPFE simulations of the laser shock compression
{\it in situ} XRD experiment of Wehrenberg {\it et al } on [110] fibre
textured tantalum, using a dislocation-based model of slip kinetics based closely on
the LMS isotropic continuum strength model of Barton {\it et al }
\cite{doi:10.1063/1.3553718}.
By extending the model to include a dislocation nucleation term, we find excellent agreement with the experimental data all the way down to the lowest shock pressures used in the experiment, and are able to elucidate the link between
the observed texture evolution and the dynamics of the underlying 
micromechanical processes of plastic deformation.  Importantly, we find that the degree of lattice rotation as a function of shock pressure observed by Wehrenberg {\it et al } 
constrains the rate of nucleation processes assumed in the model.


Our constitutive modelling approach (described in detail in the supplemental
material), includes non-linear anisotropic thermoelasticity \cite{BECKER20041983} 
and a multiscale single-crystal viscoplastic formulation based closely on the 
LMS isotropic model.
The model makes uses of the multiplicative elastoplastic decomposition of the total 
deformation gradient tensor, $\mathbf{F}=\mathbf{F}_{e}\mathbf{F}_{p}$,
where $\mathbf{F}_{e}$ and $\mathbf{F}_{p}$ are the elastic and plastic deformation
gradient tensors respectively.
Kinematics is described by the plastic component of the velocity gradient, 
which takes the form 
\cite{BRONKHORST20072351,BRONKHORST,KALIDINDI1992537,WHITEMAN201970},
\begin{equation}
\textbf{L}_{\text{p}} = 
\dot{\mathbf{F}}_{\text{p}} \mathbf{F}^{-1}_{\text{p}} = 
\sum^{12}_{\alpha=1}\dot{\gamma}^{\alpha}\textbf{m}^{\alpha}_{0}\otimes
\textbf{n}^{\alpha}_{0} \quad ,
\end{equation}
where $\dot{\gamma}^{\alpha}$ is the slip rate on slip-system $\alpha$.

Crystallographic slip is assumed to occur on 12 slip systems $\alpha$ \cite{doi:10.1179/1743280412Y.0000000015}
defined by the set of slip directions $\{\mathbf{m}_{0}^{\alpha}\}=\langle111\rangle$ 
and slip planes of type $\{112\}$ containing those directions, defined by
slip-plane normals $\{\mathbf{n}_{0}^{\alpha}\}$. 
As with the LMS model, slip kinetics is based on the Orowan equation
\begin{equation}
\dot{\gamma}^{\alpha}=\rho^{\alpha}b^{\alpha}v^{\alpha} \quad ,
\end{equation}
where $b^{\alpha}$ is the magnitude of the Burgers vector, $\rho^{\alpha}$ is the 
mobile dislocation density, and $v^{\alpha}$ is the average velocity
of mobile dislocations on slip system $\alpha$.
Some of the strain-rate and temperature dependence of the viscoplastic 
model comes from the dislocation mobility law -- the relationship between
dislocation velocity $v^{\alpha}(\tau_{\text{eff}}^{\alpha})$ and the 
effective resolved shear stress. 
This part of the model has been fitted to MD simulations of screw 
dislocation mobility, as in Barton {\it et al} \cite{doi:10.1063/1.3553718}.

The dislocation density evolution behaviour is governed by the ODE
\begin{equation}
\label{eq:dislocation_ODE}
\dot{\rho}^{\alpha}
=\dot{\rho}_{\text{mult}}^{\alpha}
-\dot{\rho}_{\text{ann}}^{\alpha}
+\dot{\rho}_{\text{nuc}}^{\alpha} \quad ,
\end{equation}
where the three terms on the RHS correspond to the processes of dislocation multiplication, annihilation, and nucleation respectively.  The LMS model accounts for the first two of these terms. Following the LMS model \cite{doi:10.1063/1.3553718}
\begin{equation}
\label{eq:dislocation_ODE_nonuc}
\dot{\rho}_{\text{mult}}^{\alpha}
-\dot{\rho}_{\text{ann}}^{\alpha}=
R\left(1-\frac{\rho^{\alpha}}{\rho_{\text{sat}}^{\alpha}
(\dot{\gamma}^{\alpha})}\right)\dot{\gamma}^{\alpha} \quad ,
\end{equation}
where $R$ is a material parameter and $\rho_{\text{sat}}^{\alpha}$ 
the strain-rate dependent saturation dislocation density for 
slip system $\alpha$, and is constrained by multiscale 
DD simulations from
Ref.~\cite{doi:10.1063/1.3553718} (see supplemental material). 
The first term accounts for 
dislocation density growth that occurs due to multiplication of pre-existing
dislocations (e.g. via the expansion of existing dislocation loops and
the Frank-Read mechanism), and the second term accounts for 
dislocation-dislocation annihilation. 

It is known that at the high strain-rates present in shock fronts, 
dislocations or twins can be nucleated, either from
heterogeneities such as grain-boundaries (heterogeneous nucleation) or, 
at very high strain-rates, nucleation out of the bulk (homogeneous 
nucleation).
Studies using MD simulations estimate that the threshold shock pressure for activation of homogeneous dislocation nucleation
is $\sim 65$~GPa \cite{doi:10.1063/1.3686538} in tantalum for [100]
loading. Other MD studies predict the homogeneous nucleation of deformation twins
at shock pressures beyond 40~GPa for [110] loading \cite{PhysRevB.88.134101}, but also in other loading 
directions \cite{Zepeda-Ruiz_2017,PhysRevB.88.104105}.
Heterogeneous
nucleation of twins and dislocations
may be expected to be important at lower shock pressures \cite{TRAMONTINA20149}. 
Deformation twinning has been seen in
in Wehrenberg {\it et al} \cite{Wehrenberg2017,PhysRevLett.120.265502} to become active at shock pressures beyond $\sim 25$~GPa.
Twinning in tantalum generates average plastic 
deformation along the same planes 
$\{\mathbf{n}_{0}^{\alpha}\}=\{211\}$ and same directions
$\{\mathbf{m}_{0}^{\alpha}\}=\langle111\rangle$ as slip, and therefore 
generates rotations in the host lattice in the same directions as does slip.
Twinning may be reasonably modelled as a pseudo-slip mechanism, 
contributing additively to the deformation rates on the original 
slip systems $\alpha$, an approach originally proposed in 
Ref.~\cite{HOUTTE1978591}, and now widely used.
We account for a dislocation nucleation process through the term
$\dot{\rho}_{\text{nuc}}^{\alpha}$ in equation \ref{eq:dislocation_ODE}. 
It takes an Arrhenius-like form,
\begin{equation}
\label{eq:nuc}
\dot{\rho}_{\text{nuc}}^{\alpha}=\dot{\rho}_{0}\exp\left\{ 
\frac{-g_{\text{nuc}}C_{44}b^{3}}{k_{\text{B}}T}
\left(1-\frac{\tau_{\text{eff}}^{\alpha}}{\tau_{\text{nuc,0}}
\left(\frac{C_{44}}{C_{0,44}}\right)}\right)\right\}~,
\end{equation}
where $\tau_{\text{nuc,0}}$ is a nucleation threshold
(at ambient pressure and temperature), $g_{\text{nuc}}$ is a material
parameter, $k_{\text{B}}$ is
the Boltzman constant, $T$ is temperature, $C_{44}$ and $C_{0,44}$ are
shear elastic modulii at current and ambient temperature and pressure respectively.
Although this form has been motivated by the homogeneous dislocation
nucleation model of Austin {\it et al}~\cite{AUSTIN20111}, we use this term
here as a proxy for nucleation processes generally, including twinning. 

The label `Model 1' is used here to refer to the model without this nucleation 
term (i.e. only the first two terms of equation \ref{eq:dislocation_ODE}
are used), and `Model 2' is used to refer to the full model that includes 
the nucleation term.
Model parameters are given in Table \ref{tab:table1}, and in the supplemental material.

\begin{table}[b]
\caption{\label{tab:table1}
Selected crystal plasticity model parameters (see supplemental materials
for the full model). }
\begin{ruledtabular}
\begin{tabular}{ccccccc}
Param. & Value & Units & &
Param. & Value & Units \\
\hline
$\dot{\rho}_{0}$  & $1.1\times10^{28}$\footnotemark[2] & m$^{-2}$s$^{-1}$ &       & $g_{\text{nuc}} $ & 0.14\footnotemark[3] & - \\
$\tau_{\text{nuc,0}}$ & $5.5$\footnotemark[3]        & GPa      &  & $R $ & $10^{17}$\footnotemark[1] & m$^{-2}$\\

\end{tabular}
\end{ruledtabular}
\footnotetext[1]{From Ref.~\cite{doi:10.1063/1.3553718}.}
\footnotetext[2]{From Ref.~\cite{AUSTIN20111}.}
\footnotetext[3]{Fitted to Wehrenberg {\it et al} \cite{Wehrenberg2017}
lattice rotation data.}
\end{table}


The simulation geometry used consists of a flyer plate region and a sample region,
each of dimensions
$4~\mu\text{m}\times1~\mu\text{m}\times1~\mu\text{m}$.
Zero velocity boundary conditions on the $y$ ($z$) component of velocity
at the $y$ ($z$) simulation cell boundaries are applied to simulate 
the (average) effect of
inertial confinement due to the effectively infinite extent 
of the experimental sample in those directions.
A shock wave is generated in the sample material by
imparting an initial velocity in the $+x$ direction to the `flyer plate'
material, so that it impacts the front surface of the sample material.

The regions are meshed up with a cubic finite elements of length
$10~\text{nm}$.
The sample consists of columnar
grains aligned with the loading axis $x$, with hexagonal cross sections of 
diameter $0.1~\mu\text{m}$, which is consistent with the
grain morphology
of the deposited samples used in the experiment of Wehrenberg {\it et al}
\cite{Wehrenberg2017}. Grain orientations are
sampled randomly from an orientation distribution function (ODF) representing a [110] fibre texture aligned with the $x$ axis, with a
spread in the angle between the [110] axis and the $x$ axis given by 
a Gaussian with full width at half max FWHM=$4^{\circ}$. Further details 
of the simulation setup are provided in the supplemental material.


\begin{figure}
\includegraphics[width=86mm]{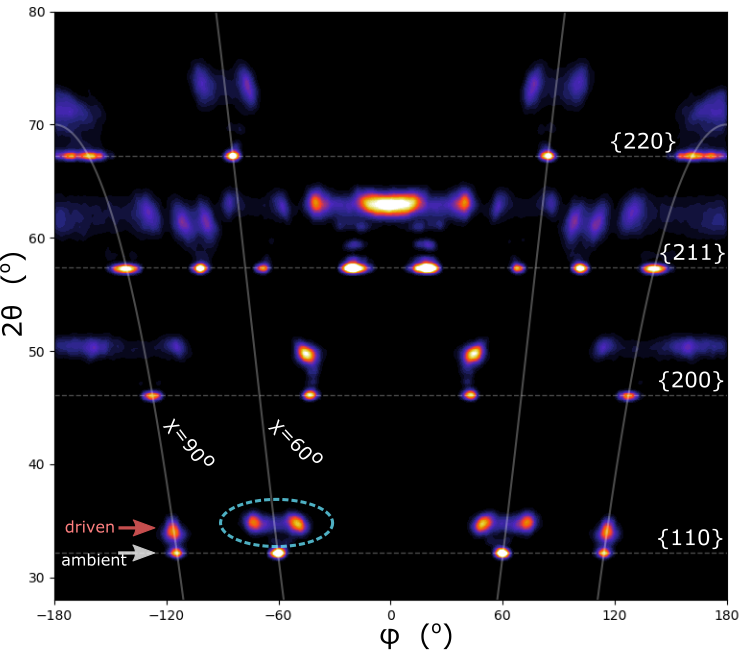}%
\caption{\label{fig:diffraction} Simulated diffraction 
(using an x-ray energy of $9.6$~k$e$V)
taken from a simulation of a 75~GPa shock, from a snapshot at time $t=0.6$~ns, 
when the shock front
has travelled approximately three quarters of the way through the sample. Both ambient and driven spots are visible. Splitting of the 
$\chi=60^{\circ}$ texture spot (e.g. those circled within the cyan
coloured dashed ring) indicate lattice rotation.}
\end{figure}

The hydrocode tracks the elastic deformation gradient field 
$\mathbf{F}_{\text{}e}(\mathbf{x})$, from which a Debye-Scherrer (DS)
diffraction pattern may be calculated for comparison with the experimental
pattern. One such simulated diffraction pattern is shown in 
Figure~\ref{fig:diffraction}, using the x-ray geometry of the
Wehrenberg {\it et al} experiment.
Details of how this was calculated are given in the
supplemental material, and in MacDonald {\it et al} \cite{doi:10.1063/1.4953028}.
Previous work on simulated
DS diffraction patterns have used assumptions
of isotropic yield strength and/or polycrystalline homogenisation based
on the Voigt or Reuss assumptions 
\cite{doi:10.1063/1.4874656,doi:10.1063/1.4953028}. 
DS diffraction derived from full-field CPFE overcomes those limitations:
it accounts for single crystal elasticity, single crystal plasticity, 
and explicit modelling of grain interactions negates the need for grain
homogenisation assumptions.

Lattice rotations were measured in Wehrenberg {\it et al} 
\cite{Wehrenberg2017} from the
azimuthal positions of the texture spots. They define angles $\chi$ 
as the angles between the sample normal 
and the normals to the lattice planes producing the diffraction spots. 
For example, for the \{110\} ring, the ambient spot at $\chi=90^{\circ}$
remains centred on the $\chi=90^{\circ}$ contour on compression
(i.e. the $\chi=90^{\circ}$ planes do not rotate) because
the rotation occurs mostly in this plane,
whereas the $\chi=60^{\circ}$ texture spot splits under compression
since the corresponding diffraction plane rotates away from the
compression axis. These features can be seen in the simulated diffraction
pattern in Figure~\ref{fig:diffraction}.

\begin{figure}
\includegraphics[width=86mm]{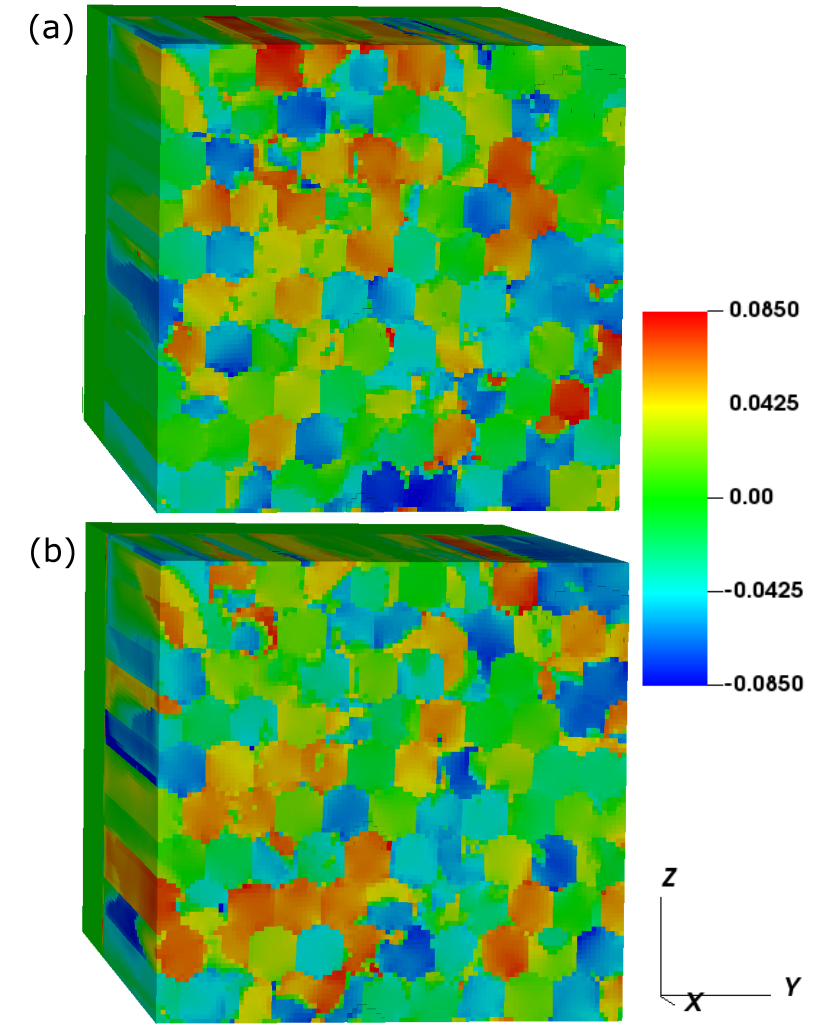}%
\caption{\label{fig:rotation_viz} 
Rotation matrix elements (a) $Q_{12}$
and (b) $Q_{13}$, for the 35~GPa shock at $t=0.6$~ns, and 
a slice at $x=1\mu$m, significantly behind the shock front,
using `Model 2'. (see supplemental material for details on rotation in this context).
These matrix elements represent reasonable approximations to the 
lattice rotation (in rads) about the $z$ and $y$ axes respectively. 
Morphology of domains of similar rotation correlate with the grain scale, though significant heterogeneity within grains too.}
\end{figure}

Figure \ref{fig:rotation_viz} shows a visualisation of lattice rotation 
\footnotemark[4] behind the shock front for a 35~GPa shock.
The simulations predict
that the morphology of domains of similar rotation correlate 
with the grains scale, there is significant heterogeneity within grains,
and some grains exhibit regions that rotate in opposing directions.
\footnotetext[4]{The rotation matrix 
whose components are plotted in the figure is
that derived from polar decomposition of $\mathbf{F}_e$.}

\begin{figure}
\includegraphics[width=86mm]{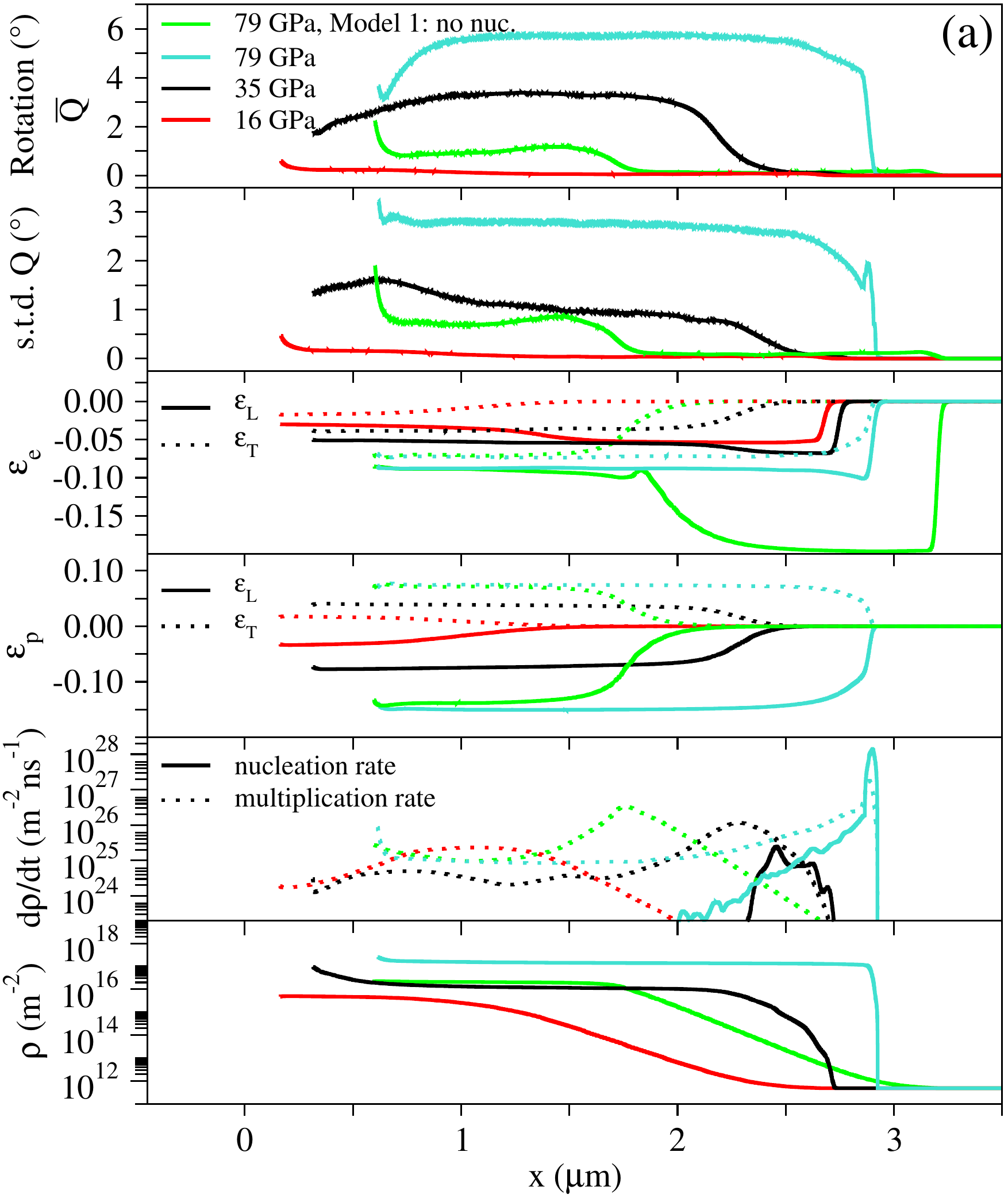}%
\newline
\includegraphics[width=86mm]{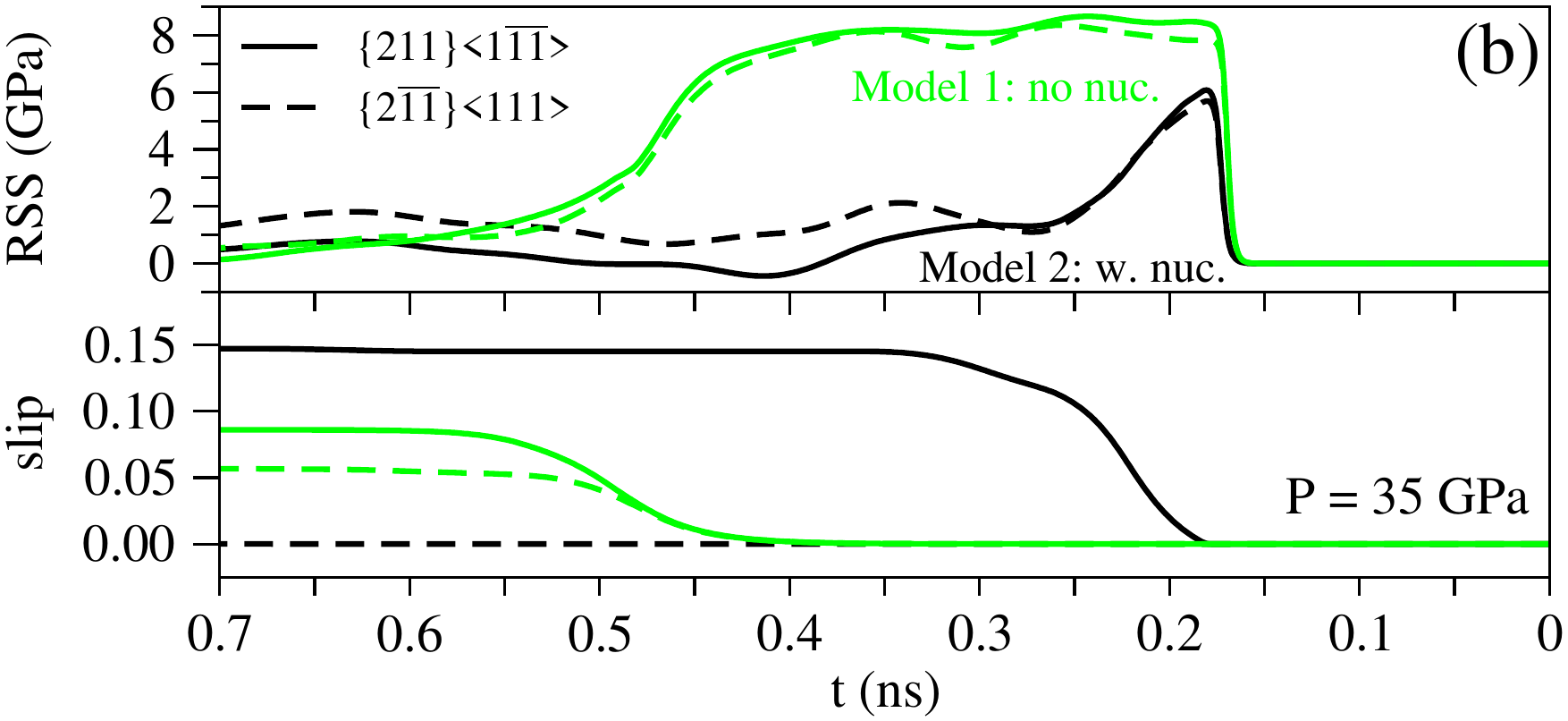}%
\caption{\label{fig:profiles} (a) Profiles of various quantities
along the loading direction ($x$) at time $t=0.6$~ns,
averaged over the lateral directions $y$ and $z$, taken from
simulations using `Model 2', unless otherwise stated. The average lattice rotation ($\bar{Q}$), the standard deviation of this lattice rotation s.t.d$(Q)$, the elastic and plastic strains $\varepsilon_e$ and
$\varepsilon_{p}$, dislocation growth rates $\text{d}\rho/\text{d}t$, 
and the dislocation density $\rho$.
(b) timeseries of the resolved shear stresses $\tau^{\alpha}$ 
and slip-system activity $\gamma^{\alpha}$ 
for the two dominant slip systems, from the $P=$35~GPa simulations,
from a material point located at $x=1~\mu$m.
}
\end{figure}

The degree of lattice rotation behind the shock front is distributed 
around a non-zero central value with an approximately Gaussian spread (see supplemental material). 
This distribution is the reason the $\chi=60^{\circ}$ spots split
in $\phi$, rather than merely becoming broader.
Figure \ref{fig:profiles}(a) shows profiles along the loading axis
($x$) of the rotation, the standard deviation of the rotation, and other 
quantities, averaged over the lateral directions $y$ and 
$z$, taken at time $t=0.6$~ns. 

\begin{figure}
\includegraphics[width=86mm]{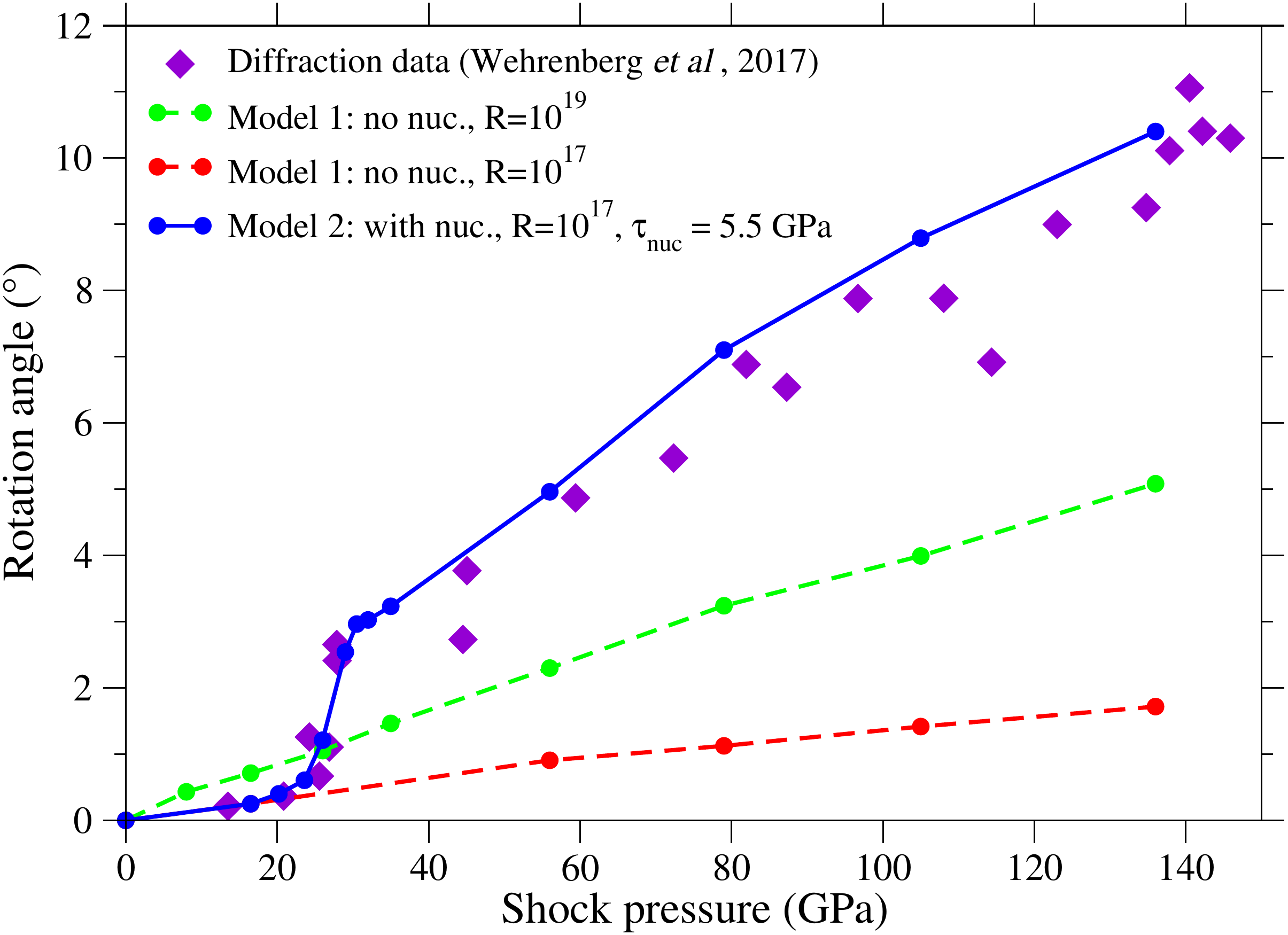}%
\caption{\label{fig:rotations} Lattice rotation from the x-ray diffraction
experiment and from CPFE modelling. Model 2 
is able to capture the low and high
pressure regimes. Model 1 is based on the LMS model and accounts for dislocation multiplication and annihilation, with multiplication rate controlled by parameter $R$. Model 2 additionally incorporates a nucleation process, with nucleation threshold $\tau_{\text{nuc}}=5.5$~GPa.}
\end{figure}

The amount of lattice rotation increases with shock pressure, seen
in Figure \ref{fig:rotations}. Two dynamical regimes can be identified,
a regime at shock pressures greater than $\sim27$~GPa, where
lattice rotations are large, and
a regime at low shock pressure, where lattice rotations are small,
even compared with the expected trend extrapolated from the higher 
pressure data. The transition between the two regimes is marked
by the onset of deformation twinning in the Wehrenberg {\it et al}
\cite{Wehrenberg2017} data, and by the onset of nucleation 
in our model. 
In our modelling using Model 2 (the model that includes nucleation),
the slip system resolved shear stress (RSS) in the shock front 
never exceeds the threshold stress for nucleation 
($\tau_{\text{nuc},0}=5.5$~GPa -- tuned to match
this data) in the low shock-pressure regime. At larger shock-pressures, nucleation becomes active.
This is confirmed in Figure \ref{fig:profiles}(a), where the dislocation
growth rates due to multiplication from pre-existing 
dislocations $\dot{\rho}_{\text{mult}}$, and
due to nucleation $\dot{\rho}_{\text{nuc}}$, are plotted: the 16~GPa shock shows no nucleation,
and as a result, the plastic and elastic strains $\varepsilon_p$ and 
$\varepsilon_e$ take longer to relax than the 35~GPa and 75~GPa shocks.

In order to investigate the effect of nucleation processes on lattice
rotation, we use `Model 1' (no nucleation) and 
`Model 2' (with nucleation) across the range of shock-pressures.
Figure~\ref{fig:rotations} shows that without nucleation, lattice rotation
remains very low across the range of shock-pressures (dashed red line).
Increasing the dislocation multiplication rate $R$ by a factor of 100 can
increase the amount the crystal rotates (dashed green line). However the
original value of $R$ has been constrained by multiscale modelling 
~\cite{doi:10.1063/1.3553718} so an increase in two orders of magnitude
cannot be justified. In contrast, `Model 2' is able to match the rotations
in the data, including the two regimes at low and high shock-pressure.

A single-slip model proposed in Wehrenberg
{\it et al} \cite{Wehrenberg2017} related the amount of lattice 
rotation to the amount of plastic 
strain (although as noted above, that simple model is not strictly valid under uniaxial conditions \cite{Heighway2021}). However, the differences in amount of rotation observed here between
Model 2 and Model 1 can not be explained by differing amounts of plastic 
strain: Figure~\ref{fig:profiles}(a) shows data for the 75~GPa 
shock with both Model 1 and Model 2, and shows that both models predict very
similar levels of plastic strain far behind the shock-front (though
the relaxation time is longer for Model 1). Despite having similar plastic 
strains, they have very different rotations.

The reasons for the discrepancy are elucidated by 
Figure~\ref{fig:profiles}(b) which shows 
the slip system resolved shear stress and slip-system activity 
at a material point located in the centre of a grain cross-section 
at position $x=1~\mu$m. The total amount
of slip predicted by Model 1 and Model 2 are similar, however Model 1
shares this slip relatively evenly between the two dominant slip systems
($\{211\}\langle1\bar{1}\bar{1}\rangle$ and  $\{2\bar{1}\bar{1}\}\langle111\rangle$),
whereas Model 2 distributes almost all of this slip to only
one of the slip systems. The amount of rotation is related to the 
discrepancy in slip between the two systems, since slip on those
systems generates rotation in mutually opposing directions.
During the elastic phase of the loading,
there is a very small discrepancy in resolved shear stress between the two 
systems, resulting from two main sources: (1) elastic interactions at 
boundaries between neighbouring grains~\cite{PhysRevMaterials.3.083602}, and 
(2) perturbations in the initial alignments the grains from perfect [110] alignment. 
With Model 2, the plastic relaxation rate near the nucleation
threshold is a very rapidly varying function of the resolved 
shear stress. 
Once the nucleation threshold is reached by 
the first slip system, plastic relaxation occurs rapidly, which
relieves the shear stress on {\it both}
slip systems, so that the second slip system never reaches the nucleation 
threshold. Thereafter, slip is much easier on the first slip system because
it has many more dislocations than the second.
With Model 1, the plastic relaxation rate at the shock front is a much 
more slowly varying function of the resolved shear stress, 
resulting in a more even distribution of defect generation
and slip between the two slip systems, 
better reflecting their initially similar RSS.

This model provides for the first time a unified picture that explains the phenomenology of the Wehrenberg {\it et al} experiment in both the small- and large- rotation regimes, overcoming the limitations of
previous modelling: the single-slip Schmid model, and MD simulations (which cannot account for realistic microstructures or non-overdriven shocks, due to computational resource limitations). We have also shown
that {\it in situ} texture evolution 
(measured using XRD) can be constraining of plastic deformation kinetics models.

The nucleation threshold of $\tau_{\text{nuc},0}=5.5$~GPa predicted here differs from the heterogeneous twin nucleation threshold at room temperature and zero pressure of $\sim1.9$~GPa predicted from constant strain rate MD simulations of single crystal Tantalum loaded in the [100] direction \cite{Zepeda-Ruiz_2017}, and from the a homogeneous twin nucleation threshold of $\sim8.2$~GPa implied by their defect free simulations. It is possible that the discrepancy between these predictions and our estimate is due to inaccuracy in the MD simulation (due to the noted inaccuracies of traditional classical interatomic potentials). It is also possible that the discrepancy is due to non-Schmid effects, which would result in the CRSS being dependent on loading axis, as it is for BCC slip \cite{doi:10.1063/5.0011708}. It is also the case that our estimate of 5.5~GPa can be influenced by other model uncertainties: the transition shock pressure in our model is determined by the balance between the rate of multiplication and the rate of nucleation. For example, we find that we are able to obtain consistency with the XRD data with a lower nucleation threshold of $\tau_{\text{nuc},0}=3.5$~GPa and a higher multiplication rate parameter of $R=10^{18}$~m$^{-2}$ (see Supplemental Material).

It is not unfeasible that the multiplication rate used in this work (parametrized by $R=10^{17}$~m$^{-2}$, derived from DD modelling) represents an underestimate of the dislocation density growth rate in the weak shock regime: modelling of plate-impact experiments in both fcc aluminium \cite{doi:10.1063/1.5008280} and bcc tantalum \cite{doi:10.1063/1.5110232} have suggested that heterogeneous dislocation nucleation processes (i.e. the nucleation of dislocations from defects such as grain boundaries and other stress concentrators), may be additionally required in order to rationalize the observed wave profile evolution. A critical resolved shear stress of 1.48~GPa for spontaneous heterogeneous dislocation nucleation in Tantalum has been suggested. The mechanical twin generation rate may also depend on initial material microstructure: MD simulation of nano-polycrystalline fibre-textured tantalum have shown a tendency for twins to nucleate from grain boundaries \cite{PhysRevMaterials.3.083602}. 


In conclusion, an understanding of the mechanisms and kinetics of plastic
deformation during high-rate loading has long been sought. 
The pioneering {\it in situ} x-ray 
diffraction experiment of Wehrenberg {\it et al} \cite{Wehrenberg2017}
at the LCLS XFEL facility, measured grain rotation as a function of pressure resulting from 
slip and twinning in [110] fibre textured tantalum. We have performed 
dislocation-based crystal plasticity finite element (CPFE) 
modelling of these experiments, from which we have generated artificial
diffraction patterns which reproduce many of the salient features
of the experimental patterns. 
We have found that the lattice rotation data from the experiment 
can well be represented by a model
that accounts for both the dislocation multiplication from 
pre-existing dislocations (based closely on the Livermore Multiscale model) and an additional dislocation nucleation term
which activates when the resolved shear stress exceeds a critical
theshold of 5.5~GPa. We discuss the sensitivity of this estimate to other plausible model uncertainties.
The degree of lattice rotation is found to be linked to the kinetics
of the processes of defect growth. When nucleation is active, 
defect growth kinetics are an extremely 
rapidly varying function of
the resolved shear stress. Very small discrepancies in 
resolved shear stress
between slip systems are strongly rewarded, leading to slip that is
concentrated on only one of the available slip systems, which results in 
large rotations. At lower shock pressures, when nucleation is not active,
the kinetics of slip are a more slowly varying function of the resolved
shear stress leading to a more equitable distribution of slips on the 
two favourably oriented slip systems, and therefore a mutual cancellation of rotations 
resulting in small net rotations. 
The data of Wehrenberg {\it et al} are thus found to be constraining of the
rate equations for slip kinetics during a shock, and the excellent agreement of the diffraction data with these CPFE simulations demonstrates how multiscale models can inform our understanding of detailed lattice-level plasticity at ultra-high strain-rates.

P.G.H. and J.S.W.\ gratefully acknowledge support from AWE via the Oxford Centre for High Energy Density Science (OxCHEDS). J.S.W. is also grateful for support from EPSRC under grant number EP/S025065/1.

UK Ministry of Defence \copyright Crown Owned Copyright 2021/AWE

\end{document}


\title{Supplemental Material: Crystal plasticity finite element simulation of laser shock-compression with \textit{in situ} 
x-ray diffraction}


\author{P. Avraam}
\email[]{philip.avraam@awe.co.uk}
\affiliation{Materials Physics Group, AWE, Aldermaston, Reading RG7 4PR, 
United Kingdom}
\author{D. McGonegle}
\affiliation{Materials Physics Group, AWE, Aldermaston, Reading RG7 4PR, 
United Kingdom}
\author{P. G. Heighway}
\affiliation{Department of Physics, Clarendon Laboratory, University of 
Oxford, Parks Road, Oxford OX1 3PU, United Kingdom}
\author{C. E. Wehrenberg}
\affiliation{Lawrence Livermore National Laboratory, Livermore, California 94550, USA}
\author{E. Floyd}
\affiliation{Materials Physics Group, AWE, Aldermaston, Reading RG7 4PR, 
United Kingdom}
\author{A. Comley}
\affiliation{Materials Physics Group, AWE, Aldermaston, Reading RG7 4PR, 
United Kingdom}
\author{J. M. Foster}
\affiliation{Materials Physics Group, AWE, Aldermaston, Reading RG7 4PR, 
United Kingdom}
\author{J. Turner}
\affiliation{Materials Physics Group, AWE, Aldermaston, Reading RG7 4PR, 
United Kingdom}
\author{S. Case}
\affiliation{Materials Physics Group, AWE, Aldermaston, Reading RG7 4PR, 
United Kingdom}
\author{J. S. Wark}
\affiliation{Department of Physics, Clarendon Laboratory, University of 
Oxford, Parks Road, Oxford OX1 3PU, United Kingdom}


\date{\today}

\begin{abstract}
\end{abstract}


\maketitle

\section{Crystal Plasticity model description}

Our constitutive modelling approach uses the common hydrocode 
assumption of fully decoupled hydrostatic and deviatoric response, a 
common assumption for cubic crystals \cite{BECKER20041983}.
A non-linear thermoelastic treatment is used for the hydrostatic 
part of the distortion (via an equation-of-state model). The deviatoric distortion is
treated with single crystal linear elasticity, with linear pressure and 
temperature dependence of the independent elastic constants, 
kinematics based on the work of Bronkhorst {\it et al} \cite{BRONKHORST20072351,BRONKHORST,KALIDINDI1992537,WHITEMAN201970}, 
and a dislocation-based 
slip kinetics model based closely on the
viscoplastic isotropic multiscale model of Barton {\it et al }
\cite{doi:10.1063/1.3553718}.

Elastic constants used for the deviatoric part of the distortion are given by,
\begin{equation}
C_{ij}=C_{0,ij}+m_{ij}T+\chi_{ij}P
\end{equation}
where $C_{0ij}$ are the elastic modulii at temperature $T=0$ K and a 
pressure $P=0$. $m_{ij}$ and $\chi_{ij}$ are material constants, parametrised
against quantum mechanical simulations \cite{PhysRevB.57.10340}
and experimental data \cite{Simmons_1971}.

Crystallographic slip is assumed to occur on 12 slip systems $\alpha$ 
defined by the set of slip directions $\{\mathbf{n}_{0}^{\alpha}\}=\langle111\rangle$ 
and slip planes of type $\{112\}$ containing those directions, defined by
slip-plane normals $\{\mathbf{m}_{0}^{\alpha}\}$. 
The total deformation gradient is decomposed multiplicatively into 
elastic and plastic parts $\mathbf{F}=\mathbf{F}_{e}\mathbf{F}_{p}$.
The kinematics of plastic slip is described by the velocity gradient, 
which takes the form,
\begin{equation}
\textbf{L}_{\text{p}} = 
\dot{\mathbf{F}}_{\text{p}} \mathbf{F}^{-1}_{\text{p}} = 
\sum^{12}_{\alpha=1}\dot{\gamma}^{\alpha}\textbf{m}^{\alpha}_{0}\otimes
\textbf{n}^{\alpha}_{0}
\end{equation}
where $\dot{\gamma}^{\alpha}$ is the slip rate on slip-system $\alpha$.

Slip kinetics is based on the Orowan equation,
\begin{equation}
\dot{\gamma}^{\alpha}=\rho^{\alpha}b^{\alpha}v^{\alpha}
\end{equation}
where $b^{\alpha}$ is the magnitude of the Burgers vector, $\rho^{\alpha}$ is the 
mobile dislocation density, and $v^{\alpha}$ is the average velocity
of mobile dislocations on slip system $\alpha$.

Some of the strain-rate and temperature dependence of the viscoplastic 
model comes from the dislocation mobility law -- the relationship between
dislocation velocity $v^{\alpha}$ and the resolved shear stress. 
This part of the model has been fitted to molecular dynamics simulations of screw dislocation mobility, as in Barton {\it et al} \cite{doi:10.1063/1.3553718}.
We generate this data using the MD interatomic potential from 
Ref. \cite{THOMPSON2015316}.
The functional form that is used to represent the MD data 
is constructed as a 
harmonic average of two terms, a `thermal activation' (T) and 
`drag term' (D), 
\begin{equation}
v^{\alpha}=\left(\left(v_{\text{T}}^{\alpha}\right)^{-1}+
\left(v_{\text{D}}^{\alpha}\right)^{-1}\right)^{-1}~.
\end{equation}
The thermal activation term (relevant at lower shear stress)
takes an Arrhenius-type form,
\begin{equation}
v_{\text{T}}^{\alpha}=\frac{v_{0}}{\exp\left[\frac{\Delta 
G_{0}}{kT}\left\langle 1-\left\langle 
\frac{\tau_{\text{eff}}^{\alpha}}{\tau_{0}\left(\frac{C_{44}}{C_{0,44}}
\right)}\right\rangle ^{p}\right\rangle ^{q}\right]-1}
\end{equation}
where $\tau_{\text{eff}}^{\alpha}$ is 
the effective resolved shear stress on system $\alpha$,
defined in Equation \ref{eq:tau_eff} below, 
$k$ is the Boltzmann factor and $\tau_{0}$ (Peierls stress), 
$\Delta G_{0}$ (activation enthalpy),$p$ and $q$ (Peierls barrier shape 
parameters) are fitted material constants.
The braces $\langle\rangle$ indicate that $\langle x\rangle=x$
for $x>0$, and $\langle x\rangle=0$ for $x<0$. 
The `drag' term (relevant at large resolved shear stresses) takes the form,
\begin{equation}
v_{\text{D}}^{\alpha}=v_{\text{off}}+\left(v_{\text{s}}-v_{\text{off}}
\right)\left(\frac{2}{\pi}\right)\arctan\left(\left\langle\frac{\pi}{2}
\frac{\tau_{\text{eff}}^{\alpha}-\tau_{\text{off}}}{\tau_{0,\text{D}}
\left(\frac{C_{44}}{C_{0,44}}\right)}\right\rangle\right)
\end{equation}
a purely phenomenological form that is close to linear when 
$\tau_{\text{eff}}$ is small \cite{doi:10.1063/1.3553718}, 
and asymptotes towards the shear wave speed
$v_{s}$ at large $\tau_{\text{eff}}$, chosen for its ability to represent 
the underlying dislocation mobility data.

The effective shear stress is defined,
\begin{equation}
\label{eq:tau_eff}
\tau_{\text{eff}}^{\alpha}=\left\langle|\tau^{\alpha}|-\lambda 
b C_{44}\sqrt{\rho_{\text{tot}}}\right\rangle
\end{equation}
where $\tau^{\alpha}$ is the resolved shear stress (RSS) on slip system
$\alpha$, $\rho_{\text{tot}}=\sum_{\alpha}\rho^{\alpha}$ is the total 
dislocation density. 
The second term is the evolving slip-system 
hardness due to forest obstacles, which is treated as athermal, 
and represented using the standard Taylor form.
Implicit in Equation~\ref{eq:tau_eff} is the assumption of isotropic
latent hardening which implies that the activity on one slip system
results in equal hardening of all potential slip systems.

The dislocation density evolution behaviour is governed by ODE,
\begin{equation}
\label{eq:dislocation_ODE}
\dot{\rho}^{\alpha}
=\dot{\rho}_{\text{mult}}^{\alpha}
-\dot{\rho}_{\text{ann}}^{\alpha}
+\dot{\rho}_{\text{nuc}}^{\alpha}
\end{equation}
where, following the model of 
Barton {\it et al} \cite{doi:10.1063/1.3553718},
\begin{equation}
\label{eq:dislocation_ODE_nonuc}
\dot{\rho}_{\text{mult}}^{\alpha}
-\dot{\rho}_{\text{ann}}^{\alpha}=
R\left(1-\frac{\rho^{\alpha}}{\rho_{\text{sat}}^{\alpha}
(\dot{\gamma}^{\alpha})}\right)\dot{\gamma}^{\alpha}
\end{equation}
where $R$ is a material parameter and $\rho_{\text{sat}}^{\alpha}$
is the saturation value of the dislocation density, defined in 
Equation \ref{eq:rho_sat} below. The first term accounts for 
dislocation growth that occurs due to multiplication of pre-existing
dislocation (e.g. the expansion of existing dislocation loops), 
which is assumed
to be proportional to $\rho^{\alpha}$. The second term accounts for 
dislocation-dislocation annihilation, which is assumed quadratic in 
$\rho^{\alpha}$. 

$\rho_{\text{sat}}^{\alpha}$ is strain-rate
dependent and was modelled using dislocation dynamics (DD) simulations in
Ref.~\cite{doi:10.1063/1.3553718}.
Following this work, we define a saturated mobile dislocation density for
each slip system as
\begin{equation}
\label{eq:rho_sat}
\rho_{\text{sat}}^{\alpha}(\dot{\gamma}^{\alpha})=\rho_{\text{sat,min}}
+\rho_{\text{sat,0}}\left(\frac{\dot{\gamma}^{\alpha}}
{\dot{\gamma}_{\text{sat}}}\right)^{\frac{1}{N_{\text{sat}}}}
\end{equation}
with parameters fitted to their simulation data.

We account for a dislocation nucleation process through the term
$\dot{\rho}_{\text{nuc}}^{\alpha}$ in equation \ref{eq:dislocation_ODE} 
It takes an Arrhenius-like form,
\begin{equation}
\label{eq:nuc}
\dot{\rho}_{\text{nuc}}^{\alpha}=\dot{\rho}_{0}\exp\left\{ 
\frac{-g_{\text{nuc}}C_{44}b^{3}}{k_{\text{B}}T}
\left(1-\frac{\tau_{\text{eff}}^{\alpha}}{\tau_{\text{nuc,0}}
\left(\frac{C_{44}}{C_{0,44}}\right)}\right)\right\}~,
\end{equation}
where $\tau_{\text{nuc,0}}$ is the nucleation threshold
(at ambient pressure and temperature). Although this form has been 
motivated by the homogeneous dislocation
nucleation model of Austin {\it et al}~\cite{AUSTIN20111}, we use this term
here as a proxy for nucleation processes generally, including twinning. 

Model parameters are given in Table \ref{tab:table1}.

\begin{table}[b]
\caption{\label{tab:table1}
Crystal plasticity model parameters. }
\begin{ruledtabular}
\begin{tabular}{ccccccc}
Param. & Value & Units & &
Param. & Value & Units \\
\hline
$C_{0,11}$ & $274.8$\footnotemark[1] & GPa    &      & $m_{11}$ & $-0.0245$ & GPaK$^{-1}$ \\
$C_{0,12}$ & $165.6$\footnotemark[1] & GPa    &      & $m_{12}$ & $-0.0118$ & GPaK$^{-1}$ \\
$C_{0,44}$ & $87.2$\footnotemark[1]  & GPa    &      & $m_{44}$ & $-0.0149$ & GPaK$^{-1}$ \\
$\chi_{11}$ & $4.22$\footnotemark[1]  & -     &      & $\chi_{12}$ & $2.78$\footnotemark[1] & - \\
$\chi_{44}$ & $0.71$\footnotemark[1]  & -     &      &  &  &  \\
\hline
$v_0$          & $250$\footnotemark[4]               & ms$^{-1}$  &      & $\tau_{0}$ & $1.05$\footnotemark[4] & GPa \\
$\Delta G_{0}$ & $0.9\times10^{-19}$\footnotemark[4] & J       &      & $p$ & $0.5$\footnotemark[4] & - \\
$q$            & $1.5$\footnotemark[4]               & -       &      & $\tau_{0,\text{D}}$ & $1.6$\footnotemark[4] & GPa \\
$\tau_{\text{off}}$  & $2.3$\footnotemark[4]         & GPa     &      & $v_{\text{s}} $ & $2000$\footnotemark[4] & ms$^{-1}$ \\
$v_{\text{off}}$  & $870$\footnotemark[4]         & ms$^{-1}$  &      & $R $ & $10^{17}$\footnotemark[2] & m$^{-2}$ \\
$b$               & $2.86\times10^{-10}$\footnotemark[1]  & m  &      & $\rho_{\text{sat,min}} $ & $5\times10^{10}$ & m$^{-2}$ \\
$\rho_{\text{sat,0}}$ & $5\times10^{19}$\footnotemark[3] & m$^{-2}$ & & $\dot{\gamma}_{\text{sat}} $ & $10^{10}$\footnotemark[3] & s$^{-1}$ \\
$N_{\text{sat}}$  & $1.3$\footnotemark[3]            & -      &       & $\lambda $ & 0.1 & - \\
\hline
$\dot{\rho}_{0}$  & $1.1\times10^{28}$\footnotemark[5] & m$^{-2}$s$^{-1}$ &       & $g_{\text{nuc}} $ & 0.14\footnotemark[6] & - \\
$\tau_{\text{nuc,0}}$ & $5.5$\footnotemark[6]        & GPa      &       &  &  &  \\

\end{tabular}
\end{ruledtabular}
\footnotetext[1]{Derived from DFT simulations in Ref.~\cite{PhysRevB.57.10340}.}
\footnotetext[2]{From Ref.~\cite{doi:10.1063/1.3553718}.}
\footnotetext[3]{Fitted to multiscale modelling data from Ref.~\cite{doi:10.1063/1.3553718}.}
\footnotetext[4]{Fitted to MD simulations using the SNAP tantalum potential.}
\footnotetext[5]{From Ref.~\cite{AUSTIN20111}.}
\footnotetext[6]{Fitted to Wehrenberg {\it et al} \cite{Wehrenberg2017}
lattice rotation data.}
\end{table}

The simulation geometry consists of a flyer plate region and a sample region,
each of dimensions
$4~\mu\text{m}\times1~\mu\text{m}\times1~\mu\text{m}$.
Zero velocity boundary conditions on the $y$ ($z$) component of velocity
at the $y$ ($z$) simulation cell boundaries are applied to simulate 
the (average) effect of
inertial confinement due to the effectively infinite extent 
of the sample in those directions. Errors due to this boundary condition
are minimal, and are further mitigated by calculating our 
integrated quantities from a middle core region of the sample region,
excluding crystal that is closer than $0.1~\mu\text{m}$ to the $y$
and $z$ boundaries. These edge effects can be mitigated by using periodic 
boundary conditions, which were unavailable to us here.
A shock wave is generated in the sample material by
imparting an initial velocity in the $+x$ direction to the `flyer plate'
material, so that it impacts the front surface of the sample material.

\begin{figure}
\includegraphics[width=86mm]{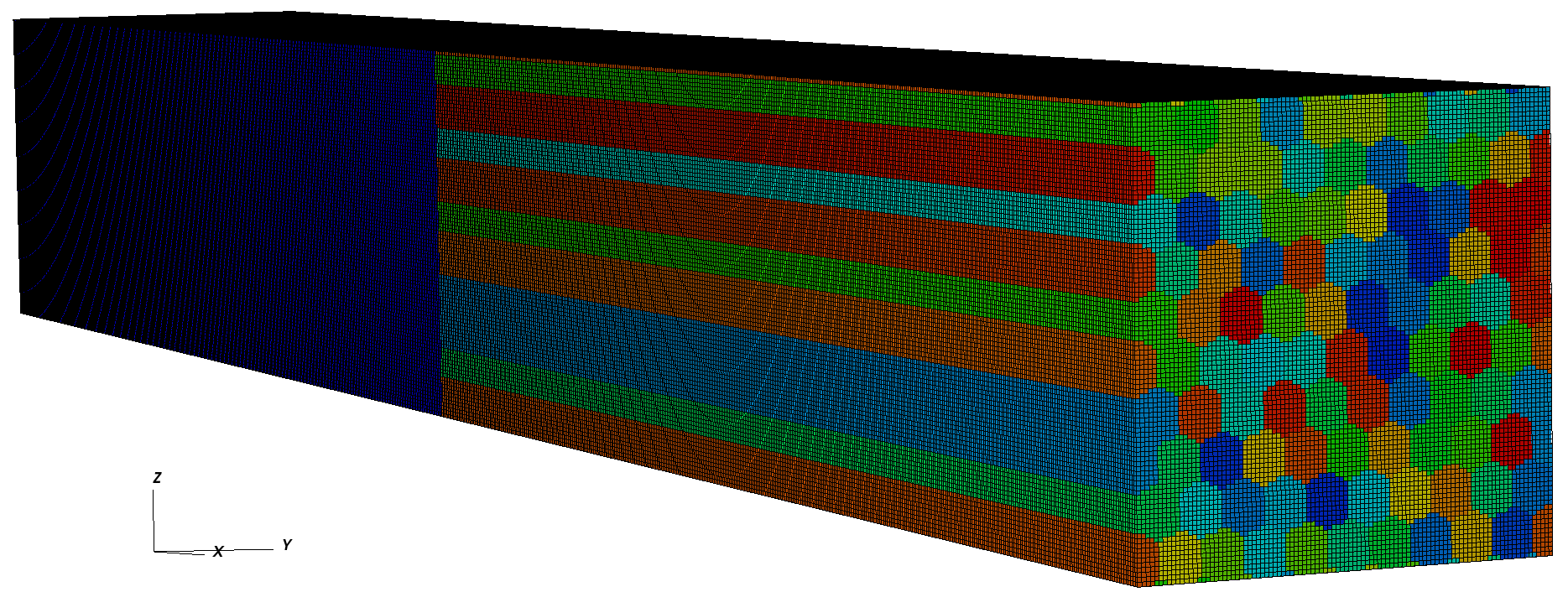}%
\caption{\label{fig:setup} Meshes and grain morphology used for all simulations.}
\end{figure}

\section{Diffraction calculation}
Debye-Scherrer diffraction patterns are calculated from these 
simulations:
The hydrocode tracks the elastic deformation gradient field 
$\mathbf{F}_{\text{}e}(\mathbf{x})$. At each material point the 
distorted lattice vectors $\mathbf{a}_{i}$ are calculated from the 
original lattice vectors $\mathbf{a_{0i}}$ using 
$\mathbf{a}_{i}=\mathbf{F}_{e}.\mathbf{a}_{0i}$. We use the method
described in MacDonald et al \cite{doi:10.1063/1.4953028} to solve the
Laue equation $\mathbf{k}-\mathbf{k}_{0}=\mathbf{G}$, where
$\mathbf{k}_{0}$ and $\mathbf{k}$ are the probe and the scattered
x-ray wavevectors, respectively, and $\mathbf{G}$ are the reciprocal
lattice vectors.

Following Ref. \cite{doi:10.1063/1.4953028}, the deviation from the
ideal Bragg angle is,
\begin{equation}
    \Delta\theta_{\text{B}} = \text{arcsin}\left(\frac{|\mathbf{G}|}{2|\mathbf{k}_{0}|}\right) + \text{arcsin}\left(
    \frac{\mathbf{k}_{0}.\mathbf{G}}{|\mathbf{k}_{0}||\mathbf{G}|}
    \right)
\end{equation}
and the relative scattering intensity can be calculated by sampling
a material rocking curve at the calculated value of 
$\Delta\theta_{\text{B}}$. In this work, we assume that the rocking
curve is a top-hat function centred around the ideal Bragg angle, i.e.
if $|\Delta\theta_{\text{B}}|<0.2^{\circ}$ then scattering occurs
with unit intensity, otherwise scattering does not occur.

The diffraction patterns are visualised by binning up the scattering
events in the $\phi$--$2\theta_{\text{B}}$ plane, where $\phi$ is the
azimuthal angle around the Debye-Scherrer (DS) ring. 

\section{Rotations}
The elastic deformation gradient $\mathbf{F}_{e}$ encodes not only
the true deformation of the crystal structure, but also local
lattice rotation. Lattice vectors $\mathbf{a}_{0i}$ are distorted according to,
\begin{equation}
    \mathbf{a}_{i}=\mathbf{F}_{e}\mathbf{a}_{0i}
\end{equation}
and lattice plane normal vectors $\hat{\mathbf{n}}_{0}$ distort as
\begin{equation}
    \hat{\mathbf{n}}=\frac{\hat{\mathbf{n}}_{0}\left(\mathbf{F}_{e}\right)^{-1}}
    {\left|\left|\hat{\mathbf{n}}_{0}\left(\mathbf{F}_{e}\right)^{-1}\right|\right|}
    \label{eq:normal}
\end{equation}

In order to disentangle the effects of rotation and distortion, 
the deformation gradient is
decomposed using the {\it polar decomposition}.
\begin{equation}
    \mathbf{F}_{e}=\mathbf{Q}_{e}\mathbf{U}_{e}
\end{equation}
where $\mathbf{U}_{e}$ is the symmetric {\it right elastic strain
tensor} and $\mathbf{Q}_{e}$ is a rotation matrix. 
Main Article Fig. 2 plots elements of the matrix $\mathbf{Q}_{e}$.

In the region undergoing plastic relaxation behind the shock front,
the rotation component typically dominates so that the orientation of lattice planes (equation~\ref{eq:normal}) may be well approximated as
\begin{equation}
    \hat{\mathbf{n}}=\hat{\mathbf{n}}_{0}\left(\mathbf{Q}_{e}\right)^{T}
    \label{eq:normal2}
\end{equation}

Lattice rotations were measured in Wehrenberg {\it et al} 
\cite{Wehrenberg2017} from the
azimuthal positions of the texture spots. They define angles $\chi$ 
as the angles between the sample normal 
and the normals to the lattice planes producing the diffraction spots. 
For example, for the \{110\} ring, the ambient spot at $\chi=90^{\circ}$
remains centred on the $\chi=90^{\circ}$ contour on compression
(i.e. the $\chi=90^{\circ}$ planes do not rotate) because
the rotation occurs mostly in this plane,
whereas the $\chi=60^{\circ}$ texture spot splits under compression
since the corresponding diffraction plane rotates away from the
compression axis. 

The degree of grain rotation is derived by assuming that all of the rotation occurs in the $\chi=90^{\circ}$ plane, and applying an appropriate geometrical transformation to recover the full rotation magnitude. The experiment is insensitive to any rotation that might
occur around the axis defined by the sample normal.

\begin{figure}
\includegraphics[width=86mm]{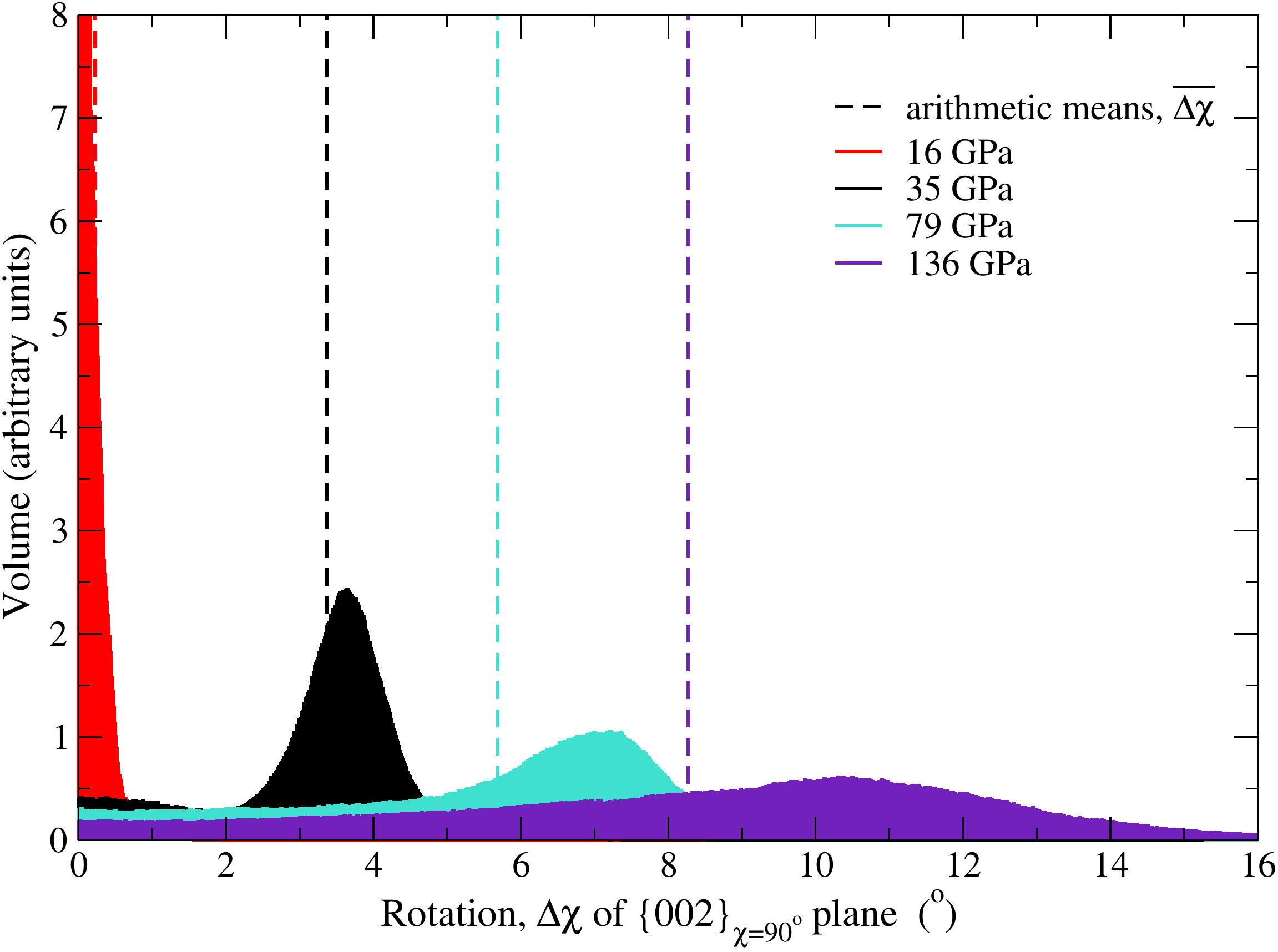}%
\caption{\label{fig:rotation_histogram} Histograms showing the
distribution of lattice rotation behind the shock front for
various shock-pressures.}
\end{figure}

For consistency with the experiment, we chose a measure of rotation
that excludes rotation around the sample normal vector: Most of the
rotation occurs in the specific \{110\} plane that makes an angle
$\chi \approx 90^\circ$ with the sample face. This plane contains the slip
plane normal of the predominant slip system, and the sample normal.
We identify the specific $\langle002\rangle$ vector that lies in 
this rotation plane,
and we calculate $\chi$ for this vector (it makes an angle of 
$\chi \approx 90^\circ$ with the sample normal) both before and after deformation, 
and we and determine 
$Q = \Delta\chi = \chi_{\text{deformed}} - \chi_{\text{ambient}}$ for this $\langle002\rangle$ vector. 

Results are plotted in Fig.~\ref{fig:rotation_histogram}.
The degree of rotation behind the shock front follows an approximately Gaussian
distribution. The peak
of the distribution differs slightly from the arithmetic mean.

In the Main Article Fig. 3, we plotted profiles of the ]
arithmetic mean lattice rotation $\bar{Q} = \bar{\Delta\chi}$.
In Main Article Fig. 4,
we calculate the lattice rotation using the peak value of
$\Delta\chi$ (rather than the arithmetic mean) which is more
consistent with the methodology used in the analysis of the
experiment, for better comparison with experimental data.

\section{Sensitivity of $\tau_{nuc}$ to $R$}

\begin{figure}
\includegraphics[width=86mm]{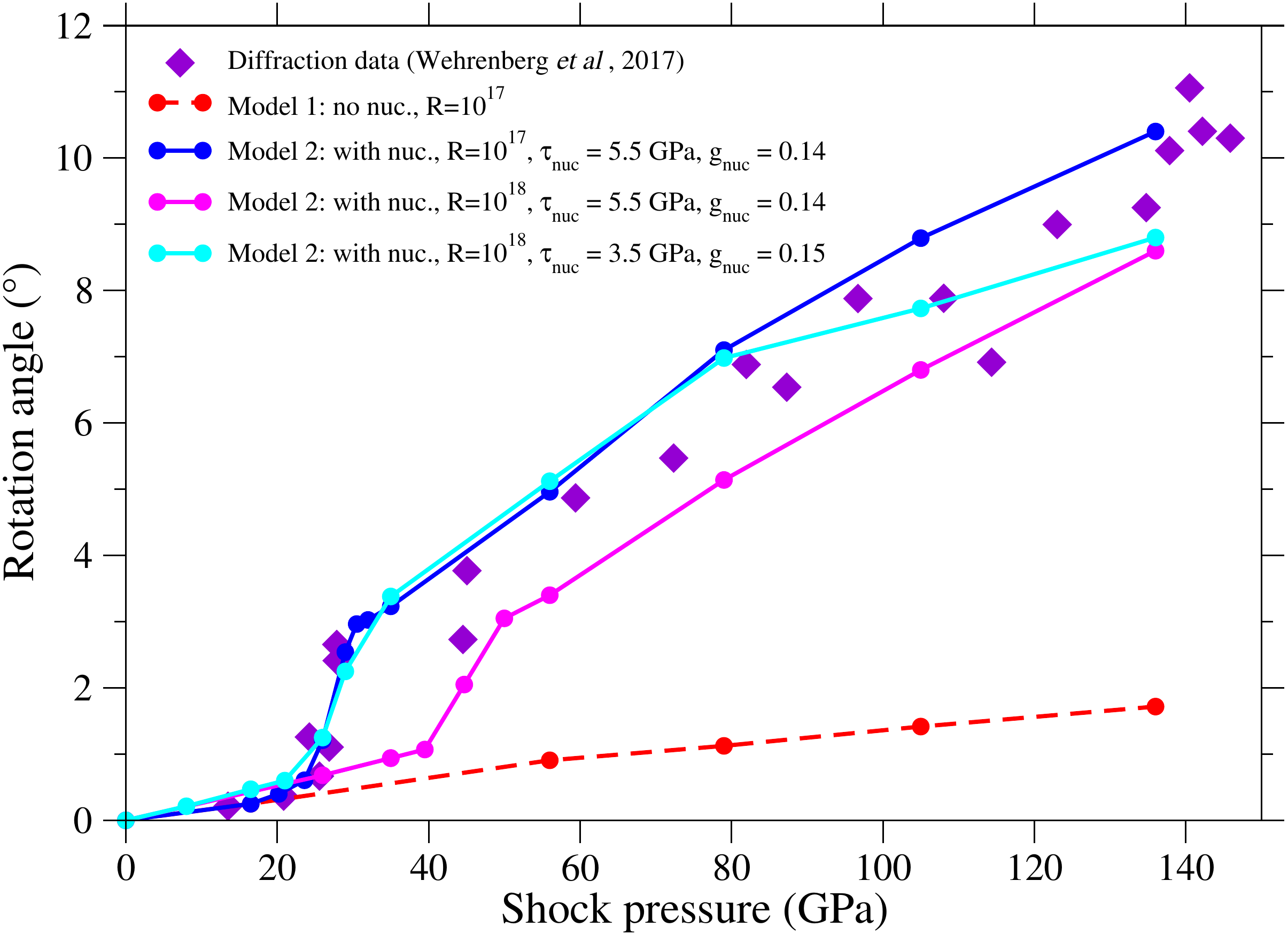}%
\caption{\label{fig:rotations2} Lattice rotation with shock pressure. The transition shock pressure is determined by the balance
between the multiplication rate and the nucleation rate. If the multiplication rate is higher, then a higher nucleation rate is needed in order to match the data}
\end{figure}

The transitional shock pressure between the regimes of low rotation and high rotation is $\approx27$~GPa. This is determined by the balance between the rates of the competing processes (multiplication and nucleation in this model). If we assume that LMS model parametrization of the multiplication process is accurate (with $R=10^{17}$~m$^{-2}$), then we find that a nucleation threshold of 5.5~GPa is required in order to match the data. This model is the blue curve in Fig.~\ref{fig:rotations2}. However, if the multiplication rate is a factor of 10 larger (i.e. $R=10^{18}$~m$^{-2}$) then the transitional shock pressure rises to $\approx45$~GPa (the multiplication process is rapid enough to keep the RSSs below the nucleation threshold at the original transition shock-pressure). Reducing the nucleation threshold to $3.5$~GPa brings the model back into consistency with the XRD data. 

It is possible that the rate of dislocation generation in the low shock-pressure low rotation regime (parametrized by $R=10^{17}$~m$^{-2}$) may be underestimated due to the lack of accounting for other mechanisms such as heterogeneous dislocation nucleation.



%